\documentclass[sigconf]{acmart}

\setcopyright{rightsretained}

\begin{document}

\copyrightyear{2019}
\acmYear{2019}
\acmConference[CECC 2019]{Central European Cybersecurity Conference}{November 14--15, 2019}{Munich, Germany}
\acmBooktitle{Central European Cybersecurity Conference (CECC 2019), November 14--15, 2019, Munich, Germany}
\acmPrice{15.00}
\acmDOI{10.1145/3360664.3360729}
\acmISBN{978-1-4503-7296-1/19/11}

\title{Approaching the Automation of Cyber Security Testing of Connected Vehicles}

\author{Stefan Marksteiner}
\email{stefan.marksteiner@avl.com}
\orcid{0000-0001-8556-1541}
\affiliation{%
  \institution{AVL List GmbH}
  \city{Graz}
  \country{Austria}
}
\author{Zhendong Ma}
\email{zhendong.ma@at.bosch.com}
\affiliation{%
  \institution{Bosch Engineering}
  \city{Vienna}
  \country{Austria}
}

\renewcommand{\shortauthors}{Marksteiner, Ma}
\renewcommand{\shorttitle}{Automated Cyber Security Testing of Connected Vehicles}

\begin{abstract}
The advancing digitalization of vehicles and automotive systems bears many advantages for creating and enhancing comfort and safety-related systems ranging from drive-by-wire, inclusion of advanced displays, 
entertainment systems up to sophisticated driving assistance and autonomous driving. It, however, also contains the inherent risk of being used for purposes that are not intended for, raging from small 
non-authorized customizations to the possibility of full-scale cyberattacks that affect several vehicles to whole fleets and vital systems such as steering and
engine control. To prevent such conditions and mitigate cybersecurity risks from affecting the safety of road traffic, testing cybersecurity must be adopted into automotive testing
at a large scale. Currently, the manual penetration testing processes cannot uphold the increasing demand due to time and cost to test complex systems. We propose an approach for an architecture that 
(semi-)automates automotive cybersecurity test, allowing for more economic testing and therefore keeping up to the rising demand induced by new vehicle functions as well as the development towards 
connected and autonomous vehicles.    
\end{abstract}

%
%



\maketitle

\section{Introduction and Motivation}
\label{sec:intro}

Automotive systems are becoming increasingly software-driven and connected. Consequently, connected vehicles must ensure cybersecurity in addition to existing stringent quality and safety requirements. Automotive systems mainly comprise of embedded devices called Electronic Control Units (ECUs) that control various parts and functions of a vehicle. The ECUs are organized into partitioned in-vehicle networks and interact with external entities through different physical and wireless interfaces such as OBDII diagnostic port, Bluetooth, WLAN and Cellular communication. With more and more connectivity-based functions and applications, automotive systems expose large attack surface shown to be vulnerable to direct and remote cyberattacks \cite{5504804,Miller2015}. 

The automotive industry as a whole starts to take cybersecurity seriously and establish secure development program and practice to address the challenges facing connected vehicles \cite{21434}. Security testing is one of the important steps in the secure development lifecycle that ensures that security designs are correctly implemented and no security vulnerabilities are left unaddressed throughout a vehicle's lifecycle. However, security testing of automotive systems is very challenging due to the complexity of software and technologies involved. Until now, it is mainly conducted by in-house or contracted penetration testers. The results of the tests often depend on the skill of human testers and the information they obtained. The thoroughness of the tests are further bound by the budget and available resource allocated. 

Even more than of traditional vehicles, security testing of connected vehicles covers a wide range of topics and technologies. Despite a plethora of suppliers in the automotive supply chain, many automotive components and subsystems share similar technologies and sometimes software components. Therefore, automated security testing has the potential to partially take over repeated and similar test scenarios and increase the efficiency of finding common vulnerabilities and weaknesses in automotive systems within vehicles from different manufacturers.    

These conditions create the need for industrialization of security testing for several reasons. First, the advancing digitalization of cars has caused a rising number of potential vulnerabilities that could not be tested manually.  
Second, this has not only led to an exponential growth of cybersecurity incidents in the last years but also to criminal hacker attacks exceeding the vulnerabilities discovered by security researchers \cite{Upstream2019}. 
And third, cost and time-to-market demands faster, more efficient testing procedures that still provide a holistic and thorough quality assurance. 
\begin{figure*}[ht!]
	\centerline{
		\scalebox{0.35}{
			\includegraphics{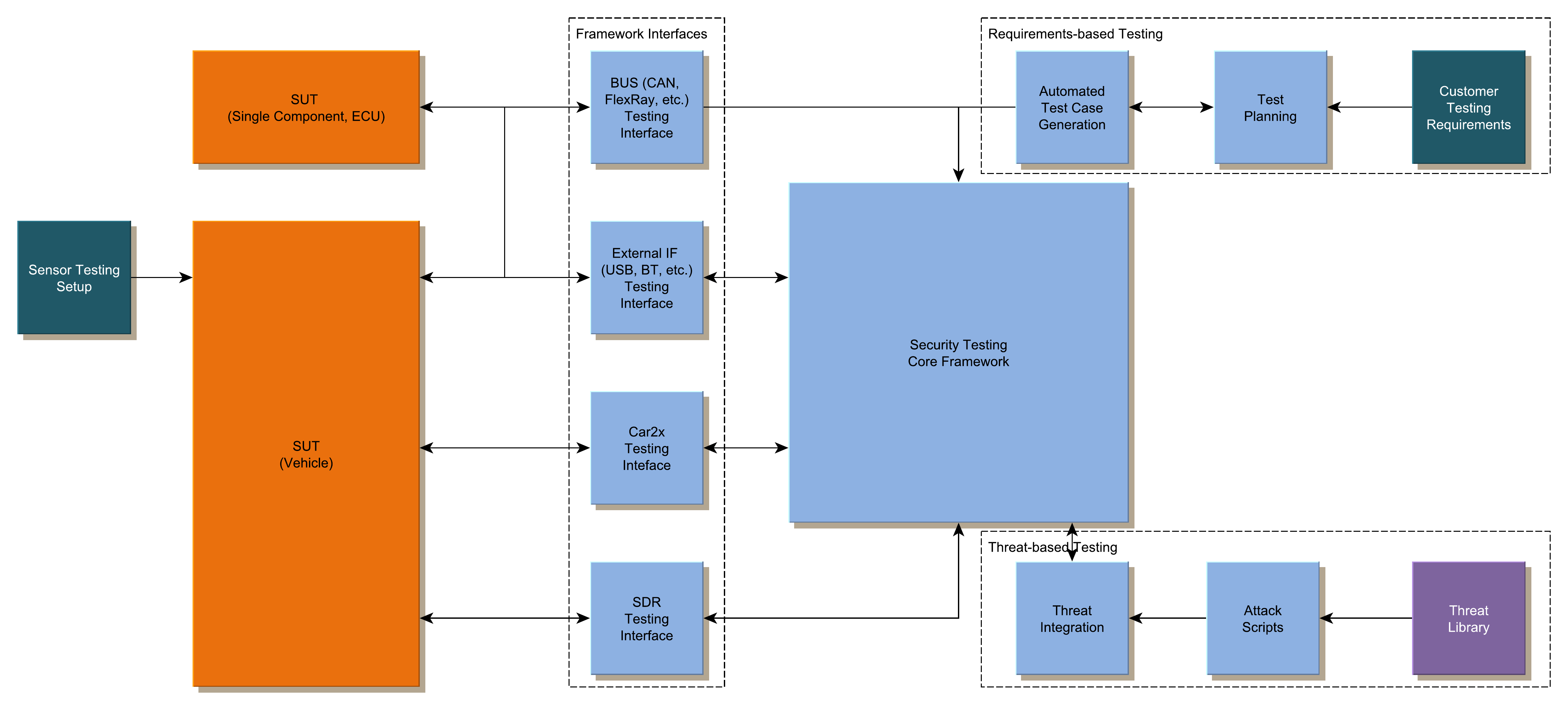}
		}
	}
\caption{Automated Testing Process}
\label{fig:arc}
\end{figure*}

Therefore, within the 
we design and develop a framework that aims at automating cyber security testing of connected vehicles. The framework provides a basis for further extensions
that allow for automating tasks that are presently carried out manually by highly specialized experts.

\section{Semi-Automated Automotive Testing}
\label{sec:AAT}
Many aspects related to product security testing are considered in the design of the framework, ranging from security test case definition, test execution and interface to automotive systems, to test 
work flow management. Our main objective is to transfer the collective knowledge of automotive security especially the existing threats and attack methods into executable test procedures leveraged by 
open source and proprietary software.    
To achieve this, we have identified three main components for a system automating security testing:
\begin{itemize}
  \item An orchestration service (or core framework);
  \item Interfaces to the system under test (SUT), which might be a full vehicle or component(s) thereof;
  \item Sources for the vulnerabilities and threats for tests.
 \end{itemize}
The orchestration service thereby works as a supervisor engine that handles the workflow consisting of generic building blocks, triggers physical implementations of these blocks and collects the results.
The building blocks derive from the vulnerability sources mentioned above. These sources contain formalized descriptions of attack and other test vectors. 
For a seamless integration of safety and security testing, as well as allowing for both white box-based \textit{functional security} testing and black box-based vulnerability testing, the sources are logically
divided in requirement-based (white box) and threat-based (black box) ones where the former follow traditional automotive testing work flows (originated from customer requirements) and the latter are collected from (possibly external) threat libraries,
e.g. Common Vulnerabilities and Exposures (CVE) databases \cite{NIST:2011}. The SUT interfaces then, a priori, define the concrete implementation of these test building blocks (which depend on the actual system)
and, during test execution, provide the actual interface between the orchestration service. Figure \ref{fig:arc} gives an overview of the process where blue boxes are for test system components, orange for the SUT, cyan for customer and
purple for external inputs.

\section{Discussion and Outlook}
\label{sec:outro}
The presented concept allows for (semi-)automating cybersecurity testing of automotive systems, by automatically deriving test cases from requirements and threat-based test sources and executing them using a work flow-based
orchestrating system. This follows a model-based testing approach. The main challenges to practically build such a system will be to 
\begin{itemize}
  \item Derive internals of the SUT;
  \item Acquire relevant test cases based on known threats;
  \item Formalize them in order to port the test cases from one SUT to another to industrialize tests;
  \item Build up standardized interface to allow for testing any given SUT with the same testing environment;
  \item Find attack paths actually exploit known threats.
\end{itemize}
These challenges can only be solved in the course of actually implemented prototypes; only then the details of the problems would fully emerge and solutions can be provided. The logical next step will therefore building
a demonstrator, which will occur in the course of 
an ongoing project.

\begin{acks}
This work was supported by the H2020-ECSEL programme of the European Commission; grant no. 783119, SECREDAS project. 
\end{acks}

\bibliographystyle{ACM-Reference-Format}
\bibliography{literature}

\end{document}